\documentclass[aps,pra, twocolumn, showpacs, superscriptaddress,preprintnumbers, amsmath, epsfig, floatfix,normalem]{revtex4-2}

\usepackage{graphicx}
\usepackage{amsmath,amssymb}
\usepackage{mathrsfs}
\usepackage[utf8]{inputenc}
\usepackage{bm}
\usepackage{physics}
\usepackage{mathtools}
\usepackage{mathptmx}
\usepackage{helvet}
\usepackage{ulem}
\usepackage{dcolumn}
\usepackage{comment}
\usepackage{ulem}
\graphicspath{ {./Figures/} }
\usepackage{textcomp}
\usepackage{graphicx,color,xcolor}

\usepackage{graphicx}
\usepackage{amsmath,amssymb}
\usepackage{xcolor}
\usepackage{color}
\usepackage[colorlinks=true,linkcolor=blue,urlcolor=blue,citecolor=blue]{hyperref}

\begin{document}

\title{Stable singular fractional skyrmion spin texture from the quantum Kelvin-Helmholtz instability} 

\author{SeungJung Huh}\thanks{Current address: Fakult\"at f\"ur Physik, Ludwig-Maximilians-Universit\"at, 80799 Munich, Germany}
\affiliation{Department of Physics, Korea Advanced Institute of Science and Technology, Daejeon 34141, Korea }
\author{Wooyoung Yun}
\affiliation{Department of Physics, Korea Advanced Institute of Science and Technology, Daejeon 34141, Korea }
\author{Gabin Yun}
\affiliation{Department of Physics, Korea Advanced Institute of Science and Technology, Daejeon 34141, Korea }
\author{Samgyu Hwang}
\affiliation{Department of Physics, Korea Advanced Institute of Science and Technology, Daejeon 34141, Korea }
\author{Kiryang Kwon}
\affiliation{Department of Physics, Korea Advanced Institute of Science and Technology, Daejeon 34141, Korea }
\author{Junhyeok Hur}
\affiliation{Department of Physics, Korea Advanced Institute of Science and Technology, Daejeon 34141, Korea }
\author{Seungho Lee}
\affiliation{Department of Physics, Korea Advanced Institute of Science and Technology, Daejeon 34141, Korea }
\author{Hiromitsu Takeuchi}\email{takeuchi@omu.ac.jp}
\affiliation{Nambu Yoichiro Institute of Theoretical and Experimental Physics}
\affiliation{Department of Physics, Osaka Metropolitan University}
\author{Se Kwon Kim}\email{sekwonkim@kaist.ac.kr}
\affiliation{Department of Physics, Korea Advanced Institute of Science and Technology, Daejeon 34141, Korea }
\author{Jae-yoon Choi}\email{jae-yoon.choi@kaist.ac.kr}
\affiliation{Department of Physics, Korea Advanced Institute of Science and Technology, Daejeon 34141, Korea }

\date{\today}
\begin{abstract}
Topology profoundly influences diverse fields of science, providing a powerful framework for classifying phases of matter and predicting nontrivial excitations, such as solitons, vortices, and skyrmions. These topological defects are typically characterized by integer numbers, called topological charges, representing the winding number in their order parameter field. The classification and prediction of topological defects, however, become challenging when singularities are included within the integration domain for calculating the topological charge. While such exotic nonlinear excitations have been proposed in the superfluid $^3$He-A phase and spinor Bose-Einstein condensate of atomic gases, experimental observation of these structures and studies of their stability have long been elusive. Here we report the observation of a singular skyrmion that goes beyond the framework of topology in a ferromagnetic superfluid. The exotic skyrmions are sustained by undergoing anomalous symmetry breaking associated with the eccentric spin singularity and carry half of the elementary charge, distinctive from conventional skyrmions or merons. By successfully realizing the universal regime of the quantum Kelvin-Helmholtz instability, we identified the eccentric fractional skyrmions, produced by emission from a magnetic domain wall and a spontaneous splitting of an integer skyrmion with spin singularities. The singular skyrmions are stable and can be observed after 2~s of hold time. Our results confirm the universality between classical and quantum Kelvin-Helmholtz instabilities and broaden our understanding on complex nonlinear dynamics of nontrivial texture beyond skyrmion in topological quantum systems. 
\end{abstract}

\maketitle

Nonlinear excitations are ubiquitous in nature and provide deep insights for understanding complex dynamics of many-body systems, classifying phases of matters, and developing information storage devices. Examples include solitons in nonlinear optics~\cite{Stegeman1999}, vortices in XY magnets and two-dimensional superfluids~\cite{Kosterlitz1973}, and skyrmions in magnetic materials~\cite{Nagaosa2013}. Often, these nonlinear excitations have been classified in terms of topological quantum number or topological charge, associated with symmetry of the underlying system~\cite{Nakahara2018}. Such topological excitations with nonzero topological numbers are considered physical substances with different dimensions in various systems, such as the baryon number associated with three-dimensional skyrmions in particle physics~\cite{Skyrme1962}, circulation flux of quantum vortices in superconductors and superfluids~\cite{Essmann1967,Williams1974}, and even bits in future memory with magnetic two-dimensional (2D) skyrmions~\cite{Nagaosa2013}.

In the current topological theory for nonlinear excitations, the topological number is computed by the integral of the pertinent field over a certain domain. Invoking the conventional topological theory requires one central assumption that the field must be continuous and, naturally, the classification of nonlinear excitations has been subjected to the same requirement. To go beyond that, we can ask the following question: can nonlinear excitations exist with singularities within the integral domain? Theoretical studies have suggested the existence of novel structures, such as the soft-core singular vortex in the orbital angular momentum of superfluid $^3$He-A phase~\cite{Parts1995,Karimaki1999}. An eccentric fractional skyrmion (EFS), akin to the singular vortex in $^3$He-A, has also been investigated in ferromagnetic spinor condensates~\cite{Takeuchi2022}. These excitations possess a singularity embedded in their spin texture, leading to unconventional internal structures that cannot be captured from the conventional viewpoint. It has been theoretically revealed that an EFS has half the spin vector's winding in a two-dimensional plane, and the spin vector discontinuity is resolved by having the anti-ferromagnetic order localized at the eccentric (or off-centered) spin singularity (Fig.~1a). Having the singular point, it spontaneously breaks the rotational symmetry, setting itself apart from conventional axis-symmetric skyrmions (Fig.~1b).

Notably, the eccentric fractional skyrmions can be generated from quantum Kelvin-Helmholtz instability (KHI) in spin-1 superfluid systems~\cite{Takeuchi2022}, providing an exciting pathway to discover exotic nonlinear excitations (Fig.~1c-e). The Kelvin-Helmholtz theory describes the instability of the interface between two flow streams in which infinitesimal perturbations of the interface grow exponentially to destroy a laminar flow~\cite{Kelvin1871,Helmholtz1868}. The interface displays a flutter-finger pattern that rolls up and enters into a turbulence regime at a later time. It can be found in various fluids of different length scales, such as oceans~\cite{Smyth2012}, the cloud bands of Jupiter~\cite{Delamere2010}, and even in the quantum fluid~\cite{Blaauwgeers2002,Volovik2002,Finne2006,Takeuchi2010,Suzuki2010,Baggaley2018,Kokubo2021}. In the KHI of immiscible multi-component superfluids, a universal (classical) regime exists~\cite{Kokubo2021}, displaying the flutter-finger pattern via the KHI. Quantized vortices can be emitted from the vortex sheet between two superfluid components from the tip of the finger in the nonlinear dynamic stage~\cite{Suzuki2010,Kokubo2021,Kokubo2022}.  The experimental studies of quantum KHI, however, have been focused on the interface of A and B phases of superfluid $^3$He~\cite{Blaauwgeers2002,Finne2006} and  scalar superfluids of bosonic and fermionic gases~\cite{Mukherjee2022,Hernandez2024}, limiting the exploration of the universal regime of KHI. This is because a vortex sheet itself is intrinsically unstable breaking up into a vortex array in the latter systems \cite{Volovik_2015}. While it can be stable in multi-component superfluids \cite{Parts1994a,Parts1994b,Eltsov2002,Hanninen2003,Kasamatsu2003,Kasamatsu2009}, nevertheless, the strong dissipation prevents the realization of the universality in the former \cite{Takeuchi2010}.

Here we report the observation of stable eccentric fractional skyrmions by realizing the universal regime of the quantum KHI in a spin-1 condensate. We adopt strongly ferromagnetic spinor condensates using $^7$Li atoms, which plays a key role in realizing the quantum KHI by having a magnetic domain wall with a sharp edge. Upon applying a magnetic gradient force, counterflow is induced along the domain wall, forming a continuous planar vortex sheet. 
Later time, the interface of immiscible superfluids is deformed to display the flutter-finger patterns. We note that the experiments are carried out in the universal regime of KHI, where the wavelength of the modulation is larger than the domain wall thickness. In the nonlinear dynamic stage, we observe magnetic droplets being emitted from the tips of the flutter fingers. We verify the EFS spin texture around the droplet by simultaneously measuring the condensate's phase and spin population. Notably, EFSs can be generated not only from the tips of the flutter fingers but also from the spontaneous splitting of an integer skyrmion with spin singularities.

\begin{figure}
\centering
\includegraphics[width=\linewidth]{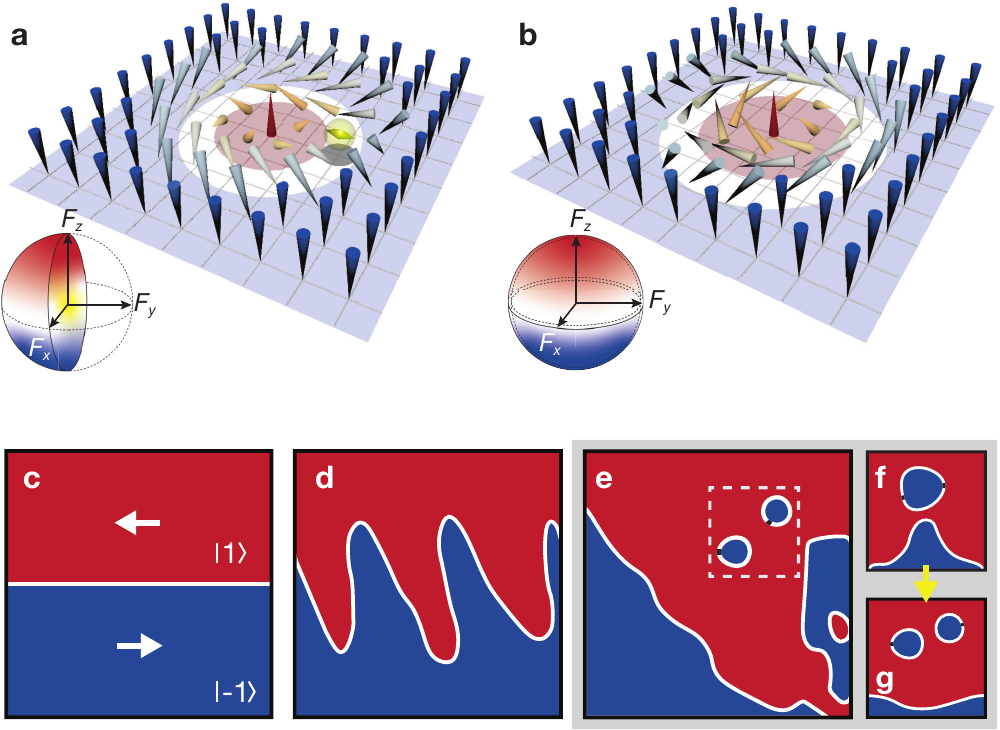}
\caption{\textbf{Eccentric fractional skyrmions from the quantum Kelvin-Helmholtz instability.} 
 \textbf{a,} Schematic illustration of the eccentric fractional skyrmion (EFS) and \textbf{b,} conventional integer skyrmion in two dimensions. The anti-ferromagnetic spin singular point (yellow double arrow) breaks the rotational symmetry of the core structure. Insets: mapping the spin vectors into the spin space, the EFS fills up half of a unit sphere, and the singular point is located at the center. The spin vectors of the integer skyrmion wind around the surface of the unit sphere.
\textbf{c-e,} The Kelvin-Helmholtz instability of spinor condensates from the counterflow between $\ket{1}$ (red) and $\ket{-1}$ (blue) spin domains.
\textbf{d,} The interface of counter-propagating flow deforms to develop a flutter-finger pattern, and later \textbf{e,} the EFS are generated from the fingertip (white dashed box). The spin singular points are marked by black dots. 
\textbf{f,} An integer skyrmion with singularities is first emitted from the fingertip \textbf{g,} and then later splits into two fractional skyrmions. 
} 
    \label{fig1}
\end{figure}

\section*{Domain wall state preparation}
The experiment began with preparing a single magnetic domain wall (DW) state of the spinor condensate in a quasi-2D optical dipole trap (Methods).
To have a deterministic creation of the initial DW state, we quench a quadratic Zeeman energy (QZE,  $q$) to an easy-axis ferromagnetic (EAF) phase ($q/h=-20$~Hz with the Planck constant $h$). The spinor condensate undergoes universal coarsening dynamics~\cite{Williamson2016,Huh2024}, and only two magnetic domains of opposite spin states ($\ket{F=1,m_z=\pm1}=\ket{\pm1}$) are left after 2 seconds of the coarsening dynamics (Fig.~2a).
From the spin-selective in-situ images~\cite{Huh2024}, we probe the longitudinal magnetization density, $F_z(\mathbf{r})$, where $\mathbf{r}=(x,y)$ is a two-dimensional coordinate, and control the axis of the DW by imposing a weak field gradient during the coarsening~(Methods and Extended Data Fig.~1).  
Since we are in the weak-easy axis regime ($|q|<\mu$ with the chemical potential $\mu$), it has a broken axisymmetric (BA) core structure with a transverse spin component (Extended Data Fig.~2)~\cite{Takeuchi2022}. To validate the experimental observations, we theoretically investigated the time evolution of the order parameter for the spin-1 Bose-Einstein condensate by numerically solving the Gross-Pitaevskii equation using our experimental parameters~(Methods).

\begin{figure*}
\centering
\includegraphics[width=0.8\linewidth]{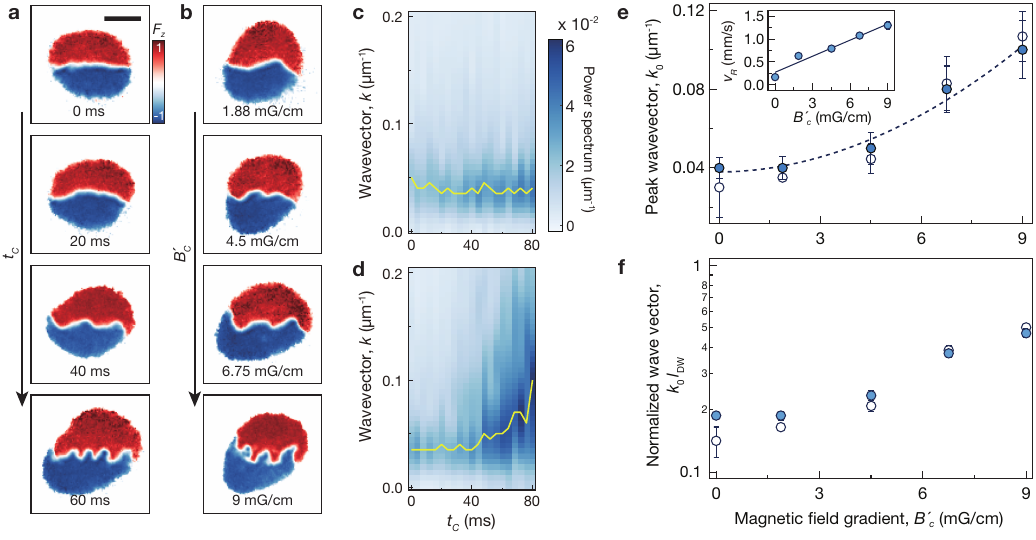}
\caption{\textbf{The quantum Kelvin-Helmholtz instability.}
 \textbf{a,} Time evolution of the longitudinal magnetization $F_z$ under $B_c'=9$~mG/cm.
Time flows along the vertical axis, developing a flutter-finger pattern in the later time.
 \textbf{b,}  Magnetization $F_z$ at $t_c=80$~ms when increasing field gradient from top to bottom.
The DW axis is set to be aligned along the horizontal axis for clarification~(Extended Data Fig.~1).
The scale bar represents 100~$\mu$m. 
\textbf{c,} Power spectrum of the interface modulation with the magnetic field gradient of 1.88~\rm{mG/cm} and \textbf{d,} 9~\rm{mG/cm}.	The solid lines indicate the peak wavevector $k_0$ for a given time $t_c$.
\textbf{e,} Peak wavevector at $t_c=80$~ms as a function of the magnetic field gradient. Closed (open) circles are the experimental data (numerical simulation), and the dashed line is the quadratic fit to data. Inset: counterflow velocity $v_R$ at $t_c=20$~ms. The solid line represents a linear fit with an offset, which is caused by nonzero spin winding along the domain wall in the initial state~(Extended Data Fig.~2).
\textbf{f,} The wavevector $k_0$ scaled by the DW thickness $l_{\rm DW}$ at $t_c=80$~ms under different field gradients. 
The data points are averaged over 20 independent experiments and error bars denote one standard deviation of the mean. 
}
\label{fig2}
\end{figure*}

\section*{Quantum Kelvin-Helmholtz instability}
A counterflow at the interface of the magnetic domain wall is introduced by applying a magnetic field gradient pulse, which imprints a spin-dependent phase gradient along the DW axis. This process generates the continuous vortex sheet, where the vorticity ${\bm \nabla}\times {\bm v}_s$ with the mass super-current velocity ${\bm v}_s$  is localized at the DW interface. As time evolves, we observe that the straight DW undergoes a deformation to display a wave pattern with a characteristic wavevector $k_0$ (Fig.~2a). At a later time, the wave modulation is pushed by the flowing spin component, and the flutter-finger pattern is well developed, which is the hallmark of the Kelvin-Helmholtz instability. The characteristic wavevector is increased when we increase the relative velocity between two spin components (Fig.~2b), consistent with the Kelvin-Helmholtz theory~\cite{Takeuchi2010, Kokubo2021}.

The quantitative study of the experimental results is followed by analyzing the power spectrum of the DW interface. Marking the two-dimensional positions of the interface $\mathbf{r}_{\rm DW}$ from $F_z(\mathbf{r}_{\rm DW})=0$, we obtain the power spectrum of the interface modulation and extract the wavevector $k_0$~(Extended Data Fig.~3). The $k_0$ remains stable in the first 20~ms of the gradient pulse, and then it rapidly rises over the hold time, indicating dynamical instability. To identify the microscopic origin of the interface instability, we investigate the dependence of the wavevector of the flutter-finger pattern on the relative velocity $v_R$, which is controlled by changing the strength of the gradient pulse $B_c'\propto v_R$. We find that the wavevector of the flutter-finger pattern follows the quadratic scaling behavior, $k_0\propto B_c'^2\propto v_R^2$ (Fig.~2e).

The quadratic dependence of $k_{\rm 0}$ on $v_R$ can be understood by considering the Bogoliubov excitation spectrum of the ripplons, quanta of the transverse vibrations of the interface. In binary superfluids, for example, the excitation {frequency with the wavevector around $k_0$} can have an imaginary value, leading to an exponential growth of the modulation amplitude ~\cite{Takeuchi2010,Kokubo2021,Takeuchi2022}. The characteristic wavevector is proportional to the square of the relative velocity~\cite{Takeuchi2010}, and similar results can be found {for our spinor condensates}~\cite{Takeuchi2022}. Although our experiments have been carried out with the BA-core domain wall state, having all three spin components in the DW core, the transverse vibration does not change the core structure, which results in keeping the spin populations almost the same during the dynamics.

Note that our experiments demonstrate the universal feature of the KHI that flutter-finger patterns are always developed when the condition $k_0l_{\rm DW}<1$ is satisfied, where $l_{\rm DW}$ is the domain wall thickness. In this regime, the DW's motion is universally described by the KH theory because the detailed structure of the DW can be neglected by the surface tension of the DW~\cite{Takeuchi2010,Kokubo2021}. In our experiment, the domain wall thickness is so small with $l_{\rm DW}=4.7~\mu{\rm m}$ due to the strong ferromagnetic spin interaction, satisfying the universality condition $k_0l_{\rm DW}\leq 0.47(2)$ (Fig.~2f). Performing the experiment in this universal regime is also crucial for clearly observing the EFS. This is because, in the opposite limit $k_0l_{\rm DW}\gg 1$, the interface modulation can develop into a sealskin or zipper pattern~\cite{Kokubo2021}. In such cases, skyrmions only with a small core size could be generated or vortices are buried in the thick interface layer, making it hard to identify the singular structure of the EFS.

\begin{figure}[t]
\centering
\includegraphics[width=0.6\linewidth]{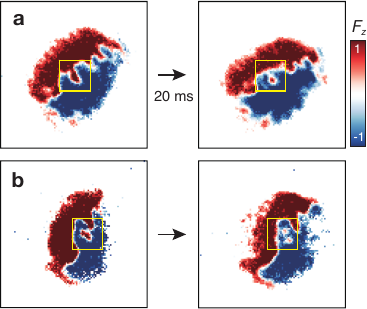}
\caption{\textbf{Nonlinear dynamics after the Kelvin-Helmholtz instability.} 
\textbf{a,} Emitting a magnetic droplet of the spin-up component to the spin-down background domain (yellow boxes), the flutter-finger pattern is collapsed.
\textbf{b,} Splitting dynamics of a single magnetic droplet into two magnetic droplets. It could indicate a spontaneous formation of two eccentric fractional skyrmions (yellow boxes). Each set of images is obtained by taking two consecutive absorption images of $\ket{\pm1}$ states with 20~ms of time interval~(Methods).  
}
\label{fig3}
\end{figure}

\section*{Generation of eccentric fractional skyrmion}

One of the distinctive features of the quantum KHI from the classical system is the generation of topological defects like quantized vortices in the nonlinear dynamic stage because the vortex circulation is quantized in the bulk spin domain as $\oint {\bm v}_s\cdot d{\bm r}=N_vh/M$ with the atomic mass $M$ and integer $N_v$. For a spin-1 system, EFSs, a new type of vortex skyrmions, have predicted to be emitted from the tip of the finger~\cite{Takeuchi2022}. Searching for the EFS, we monitor the later dynamics for a various hold time $t_h$ after turning off the gradient pulse. We capture the emission process by taking two consecutive images of the longitudinal magnetization, $F_z(\mathbf{r},t)$ and $ F_z(\mathbf{r},t+\Delta t)$, with a finite time interval $\Delta t=20$~ms (Methods). Fig.~3a shows that a magnetic droplet is detached from one of the fingertips of the spin-up component. Later, it propagates into the spin-down magnetic domain, and the finger pattern is destroyed. Furthermore, Fig.~3b displays the splitting of a single droplet into two, which might suggest the decay of an integer skyrmion into two EFSs. The numerical simulations support these observations that the spin texture around the magnetic droplet is EFS, and importantly, an integer skyrmion with spin singularities spontaneously decays into two EFSs~(Extended Data Fig.~4).

\begin{figure}[b]
\centering
\includegraphics[width=0.9\linewidth]{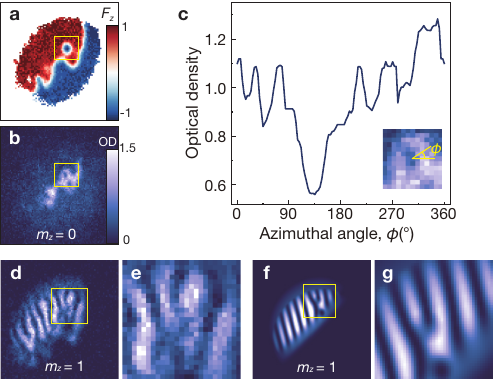}
 \caption{\textbf{Eccentric fractional skyrmion.} 
\textbf{a,} Longitudinal magnetization $F_z$ and \textbf{b,} density profile of $\ket{0}$ spin state ($n_0$) at $t_h=60$~ms. The yellow box in each image highlights the region that is likely to be the EFS.
\textbf{c,}The angular density profile of $n_0$ at the boundary of the magnetic droplet, $\mathbf{r_s}$, with  $F_z(\mathbf{r_s})=0$. The density dip indicates the off-centered anti-ferromagnetic spin singular point. 
\textbf{d,}  Matter-wave interference image of $\ket{1}$ spin states. 
The Y-shaped (two-to-one) fringe represents a phase winding of $2\pi$ around the center of the magnetic droplet (yellow box).
\textbf{e,} Enlarged image of the region at the phase singularity.
Images of a,b, and d are obtained in a single experimental run~(Methods), and highlighted regions with yellow boxes represent the same area.  
\textbf{f,} Simulated interference pattern with the eccentric fractional skyrmion \textbf{g,} and zoomed image of the central region with a phase singularity.
}
\label{fig4}
\end{figure}

Remark that the EFS spin texture has broken rotational symmetry with spin singularity beyond the current theoretical framework for topological excitations in spin-1 condensates, but still tries to satisfy the topological relationships somehow or other. For instance, the Mermin-Ho relation associates the skyrmion charge $N_s$ with the vortex winding number $N_v$~\cite{Mermin1976},
\begin{equation}
    N_s=N_v/2.
    \label{eq:MH}
\end{equation}
This relation implies the unit vortex charge in the EFS, $2N_s\simeq N_v=N_{-1}=1$ with the vortex winding number $N_m$ in the spin $\ket{m}$ state. This is because the EFS has (pseudo) topological charge $N_s \simeq1/2$, approximately evaluated by the integral of the skyrmion charge density over the domain excluding the singularity. Such a complicated structure inside the core of a topological excitation often results from the length hierarchy \cite{Salomaa1987,Takeuchi2021a,Takeuchi2021b}. In our case, there are two distinct length scales: the longitudinal-spin bending length $\xi_q \sim \hbar/\sqrt{M|q|}$, and the spin healing length $\xi_s \sim \hbar/\sqrt{M|c_s|n}~<\xi_q$, where $|c_s|n$ is the spin interaction energy (Methods). These scales characterize the thickness of the BA-core wall and the size of the singularity inside an EFS, respectively.

Likewise, the vortex winding rule~\cite{Isoshima2001} for a quantum vortex with axis-symmetric wave functions can approximately apply to our case. For example, in spin-1 condensates an axis-symmetric quantum vortex state follows the rule $N_1+N_{-1}=2N_0$. In other words, a vortex state spontaneously breaks rotational symmetry if this rule is not followed \cite{Takeuchi2021a,Takeuchi2021b}. Although the EFS breaks the rotational symmetry, it still approximately satisfies the rule: $\ket{0}$ state exhibits half winding $N_0{\simeq 1/2}$ and $\ket{1}$ state has no vortices with $N_{1}=0$. This fractional winding can be understood from the ``C"-shaped density distribution of $n_0(\mathbf{r})$ (Fig.~1a). From the in-plane spin vector distribution, the phase profile of $\ket{0}$ state has a $\pi$ winding along the ``C"-mark. This interpretation is consistent with the current velocity rule, $\mathbf{v}_1+\mathbf{v}_{-1}=2\mathbf{v}_0$ with the velocity $\mathbf{v}_m$ of the $\ket{m}$ state along the BA-core wall~\cite{Takeuchi2022}, by being approximately applied to the curved BA-core wall that wraps around the EFS. Therefore, even with the broken axial symmetry, the EFS can approximately satisfy the vortex winding rule.

To verify the emitted magnetic droplet in Fig.~3 is the EFS, we simultaneously measure density profiles of all spin components $n_{0,\pm1}(\mathbf{r})$ and the relative phase of spin $\ket{1}$ state in a single experimental run. These measurements can show the key features of the EFS that exhibit half winding of the spin vector around its core and off-centered anti-ferromagnetic spin singular point (Fig.~1a). First, the off-centered singular point can be identified by a ``C"-shaped atomic density in the spin $\ket{0}$ state around the EFS core. It is distinctive from a conventional axis-symmetric skyrmion spin texture that displays an ``O"-shaped density distribution in $n_0(\mathbf{r})$ (Fig.~1b). Imaging all three spin components of the spinor condensate, we find the ``C"-shaped density of $n_0(\mathbf{r})$ around the magnetic droplet of a spin-down component, as shown in Fig.~4. This indicates the spin singular point, $\mathbf{F}(\mathbf{r}_{\rm s})=0$. The spin singular point can also be observed in the mother DW even before the emission of the skyrmion spin texture~(Extended Data Fig.~6), which is consistent with numerical simulation~\cite{Takeuchi2022}.

Second, we probe the half winding of the spin vector around the core by utilizing a two-photon Raman matter-wave interference technique~\cite{Clade2009}. Without phase singularity, a stripe pattern can be observed, where its wavelength is determined from the momentum difference during the Raman pulses (Methods and Extended Data Fig.~5). With a vortex, however, it will show a dislocation in the stripe interference pattern because of phase singularity. The skyrmion in Fig.~1b, for example, shows a single winding $N_s=1$ of the spin vector in the two-dimensional plane around its core. From the Mermin-Ho relation {of Eg.~(\ref{eq:MH})}, the spin $\ket{-1}$ state can have a winding number of $N_v=N_{-1}=2$, i.e., $4\pi$ phase winding around the core. Because of the 4$\pi$ phase winding, the interference pattern will show a three-to-one (fork-shaped) pattern. In our experiment, we frequently observe a two-to-one (Y-shaped) pattern after the self-interference (Fig.~4d), which directly shows the $2\pi$ phase winding ($N_v=1$) around the magnetic droplet. This implies the spin vector winding number is half $N_s=1/2$ like the EFS spin texture. Note that the interference images are obtained after the partial measurement of atomic density in Fig.~4a and b so that we can confirm the essential features of the eccentric fractional skyrmion.


\section*{Long-time dynamics}
As a basic step for studying the dynamic stability of the EFS, we count the number of EFS as a function hold time $t_h$ and plot the result in Fig.~5a. The position of the EFS is marked from both the density image and the Y-shaped interference pattern. The integer skyrmion with $N_s=1$ is also found, but most of the defects are shown to be EFSs~(Fig.~5a inset), suggesting the latter is more stable than the former. This is attributed to the fact that an integer skyrmion has higher hydrodynamic energy than that of two EFSs since an integer skyrmion carries a doubly charged vortex $N_v=2$, whereas an EFS has a unity charge vortex $N_v=1$. 

\begin{figure}
\centering
\includegraphics[width=0.77\linewidth]{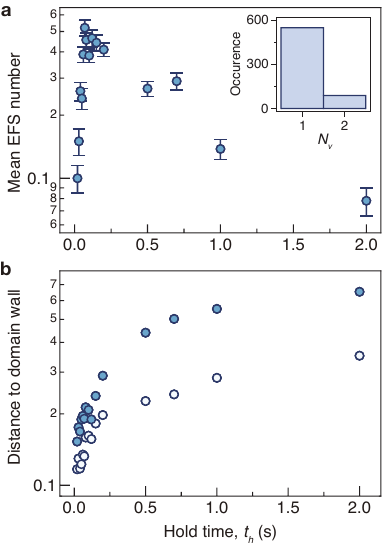}
\caption{\textbf{The birth and death of the eccentric fractional skyrmion.} 
\textbf{a,} Average number of eccentric fractional skyrmion (EFS) per image as a function of hold time $t_h$. 
The inset shows the histogram of mass vorticity measured from matter-wave interference. 
The integer vortex ($N_v=1$) corresponds to the EFS and a doubly charged vortex with $N_v=2$ could indicate the integer skyrmion.
\textbf{b,} The time evolution of the relative position of the EFS with respect to the domain wall. 
The nearest distance is normalized by the condensate radius (closed circles) and the domain length (open circles).
Each data point was obtained from 100 independent experimental realizations, and the error bars represent the one s.e.m.   
}
\label{fig5}
\end{figure}

The probability of finding the EFS increases rapidly in the first 100~ms and exponentially decreases with a decay time $\tau_{\rm EFS}=520$~ms.
We can neglect the collisional decay process between fractional skyrmions since there are at most two fractional skyrmions in the experiment and the average number of the EFS per image is below one (Fig.~5a). The two distinctive time scales can be also seen in the trajectory of the EFS (Fig.~5b). 
In the early stage, the nearest distance $l_c$ between the fractional skyrmion core and the DW rapidly increases, indicating that the EFS is generated from the domain wall boundary. Afterwards, it grows slowly on a very long time scale (570 ms). In the late time, the EFS can decay by drifting outside of the condensate like vortices in a scalar BEC~\cite{Kwon2014}. The atom loss from the microwave dressing field becomes an additional decay channel of the skyrmion spin texture. It could destroy the EFS core and decay to a singly charged vortex without any spin texture in the late time ($t_h>1$~s, see Extended Data Fig.~7).

\section*{Discussion and outlook}
Since Kelvin's seminal work on counterflow instability, the ideal realization of the Kelvin-Helmholtz instability has been hindered by the finite viscosity of ordinary fluids. 
Using ultracold atomic gases, we have achieved the first realization of the Kelvin-Helmholtz instability in an inviscid fluid and discovered eccentric fractional skyrmions that are beyond the scope of traditional topological theory. 
Our work is foundational, opening the door to the exploration of other exotic nonlinear excitations, such as defects with line singularities, and paving the way for a new framework for understanding topological defects.
Moreover, the study of counter-flow instability in multi-component superfluids promises to find an alternative route for quantum turbulence~\cite{Kobyakov2014}, complementary to driven systems~\cite{Henn2009,Navon2016,Galka2022,Hong2023}, and spin turbulence, where the ferromagnetic spin-1 condensate predicts $-7/3$ power-law Kolmogorov scaling~\cite{Fujimoto2012}.
Dynamical studies of the EFS are an exciting direction for future works that could form anomalous vortex lattice structures in spinor condensates~\cite{Schweikhard2004,Kasamatsu2004}.
More broadly, our work suggests ferromagnetic superfluids as a unique platform to realize quantum spin phenomena, that have been imagined, but unachievable in solid-state systems. 
Notably, employing the idea of skyrmion qubits~\cite{Psaroudaki2021,Xia2023}, the long lifetime of the EFS could be applicable for a storage qubit in skyrmion quantum processors. 

\section*{Acknowledgments}
We acknowledge discussions with Grigory E. Volovik, Woo Jin Kwon, Magnus O. Borgh, and Yong-il Shin.
\textbf{Fundings:} J.-y.C. is supported by the National Research Foundation of Korea (NRF) Grant under Projects No. RS-2023-00207974, RS-2023-00218998, and 2023M3K5A1094812. 
W.Y. and S.K.K. are supported by the Brain Pool Plus Program through the NRF funded by the Ministry of Science and ICT (2020H1D3A2A03099291) and the NRF grant (2021R1C1C1006273).
H.T is supported by JST, PRESTO Grant No. JPMJPR23O5, Japan, JSPS KAKENHI Grants No. JP18KK0391 and JP20H01842.

\bibliography{Ref_QHI.bib} 


\setcounter{figure}{0}
\renewcommand{\figurename}{Extended Data}
\renewcommand{\thefigure}{Fig. \arabic{figure}}

\section*{Methods}\label{sec11}
\noindent
\textbf{Experimental systems.}
Our experiments have been performed with degenerate spinor Bose-gas of $^{7}\mathrm{Li}$ atoms in a quasi-two-dimensional optical dipole trap~\cite{Huh2020}. 
The degenerate Bose gas is generated by plain evaporation cooling under a 1~G of a magnetic field along $z$-axis so that all atoms are populated to the $\ket{F=1,m_z=0}=\ket{0}$ state, called polar phase~\cite{Kawaguchi2012}. 
After its final evaporation cooling, the condensate contains $2.7\times 10^6$ atoms, and the trap frequencies of the optical potential are $(\omega_x,\omega_y,\omega_z) = 2\pi\times(7,8,680) ~\mathrm{Hz}$.
Since the chemical potential of the condensate ($\mu/h = 310~\mathrm{Hz}$) and the spin interaction energy ({$c_sn/h=-150~\rm{Hz}$} with density $n$) are smaller than the axial trap frequency, our system can be described in two dimensions. 
Moreover, the condensate fraction is over $93\%$, so thermal fluctuations like phonons and vortex pairs {nucleation} can be neglected.
Indeed, no spin excitations nor density modulations~\cite{Choi2012b} are observed along the vertical axis in this experiment.  \\

\noindent
\textbf{Magnetic domain wall.}
As a first step to study the interface instability, we prepare a single magnetic domain wall with spin $\ket{F=1,m_z=\pm1}=\ket{\pm1}$ state with its axis aligned to the $x$-direction (Extended Data Fig.~1). 
To have a larger size of the spin domain, we utilize the universal coarsening dynamics after quenching the quadratic Zeeman energy (QZE) $q$ from the polar phase to the easy-axis ferromagnetic (EAF) phase, $q/h < 0$, ~\cite{Williamson2016,Huh2024}. 
That is, in the EAF phase, the ground favors spin $\ket{\pm1}$ state, so the polar phase is dynamically unstable and generates multiple magnetic domains of $\ket{\pm1}$ state. 
When the coarsening dynamics are terminated, there are only two magnetic domains with a single domain wall. 
The axis of the domain wall is controlled by applying an additional field gradient, which exerts a spin-dependent force during the coarsening dynamics.
This method provides more atom number in the condensates than the evaporating cooling in the optical trap with $\ket{\pm1}$ state since the scattering lengths of the spin $\ket{\pm1}$ states are much smaller than that of the $\ket{0}$ state~\cite{Huh2020}.

Experimentally, we quench the QZE to the weak easy-axis ferromagnetic (EAF) phase ($q/h=-20~$Hz) by tuning the microwave frequency and apply a field gradient $B^{'}_{\rm DW}=2~\rm mG/cm$.
After 2~s of coarsening dynamics, a single well-aligned magnetic domain wall state can be prepared.
Due to the atom loss from the microwave dressing, we note that the number of atoms constituting the magnetic domain is reduced to $1.8\times10^6$.

In the weak EAF phase, the domain wall has broken-axis (BA) core~\cite{Takeuchi2022} that the spin vector continuously rotates across the wall and has a finite spin $\ket{0}$ state at the interface of the $\ket{\pm1}$ spin domains. 
This feature is directly imaged by taking the density profiles $n_m$ of the spin $\ket{m}$ component (Extended Data Fig.~2), displaying the finite spin population of the $\ket{0}$ state in between the spin $\ket{\pm1}$ state. 
Reconstructing the longitudinal magnetization, $F_z=\frac{n_1-n_{-1}}{n_1+n_0+n_{-1}}$ with the density $n_m$ of spin $\ket{m}$ state, we find the magnetization follows ${F_z(y)}=F_{z0}\tanh\left({y}/{\xi_{\rm BA}}\right)$ (Extended Data Fig.~2d).
The domain wall thickness $\xi_{\rm BA}$ is measured to be 4.7~$\mu$m. 
The BA-core domain wall state can have a long-wavelength excitation, Bloch lines, so the spin vector field is helical lying on the $xy$-plane along the domain wall and $F_x$ can have opposite signs (Extended Data Fig.~2c) as it is prepared from relaxation dynamics. 
The effect of the Bloch line on the interface instability, however, is negligible since the domain wall instability is observed in a much larger counterflow.  \\

\noindent
\textbf{Power spectrum of the domain wall interface}
For quantitative analysis of the domain wall instability, we analyze the power spectrum of the domain wall interface. 
This analysis begins with finding the set of positions $\{\textbf{r}_{\rm DW}=(x,y)\}$ corresponding to the interface of two magnetic domains, where $ F_z (\textbf{r}_{\rm DW}) = 0$ (Extended Data Fig.~3). 
We then find the linear fit function for the domain wall positions, which can be assumed to be straight when there is only rotational motion without modulation instability. 
As a measure that quantifies how the domain wall interface has deviated from the straight line, we find the height $h$ from the minimum distance between the linear function and the domain wall. 
We then evaluate the Fourier transformation of $h$ as a function of the position  $r'$ along the projected domain wall axis. 
By averaging 20 experimental realizations, we obtain the power spectrum of domain wall waves as a function of pulse duration. \\

\noindent
\textbf{Stroboscopic magnetization imaging.}
To track the domain wall dynamics, we take the two consecutive $\textit{in situ}$ magnetization imaging, separated by the finite time interval $\Delta t$, in a single experimental realization. 
The detailed imaging sequence is as follows.
We first take the longitudinal magnetization $F_z(\mathbf{r},t)$ by partially transferring atoms in each spin state into the $\lvert F=2 \rangle$ state and subsequently applying a $\lvert F=2 \rangle \rightarrow \lvert F'=3 \rangle$ resonant light. 
We set the transfer fraction to be about $20\%$, corresponding to $\sim 5\%$ reduction of the chemical potential, and it has a negligible effect on the spin dynamics within 100~ms. 
After $\Delta t = 20~\mathrm{ms}$ of the first image, we obtain the second magnetization image $F_z(\mathbf{r},t+\Delta t)$.  \\

\noindent
\textbf{Matter-wave Raman interference.}
To measure the relative phase information of trapped condensate, we adopt a matter-wave self-interference technique using a two-photon Raman transition~\cite{Clade2009}. 
It consists of two $\pi/2$ Raman pulses, separated by a finite time interval $\Delta t_{R}$, where we use different Raman laser beam pairs in each $\pi/2$ pulse.
Each pair of beams has frequency $\nu_1$ and $\nu_2$, where the frequency difference is set to have a resonant transition between two hyperfine spin states ($\lvert F=1, m_z=1 \rangle=\lvert 1, 1 \rangle$ and $\lvert F=2, m_z=2 \rangle=\lvert 2, 2 \rangle$, see Extended Data Fig.~6a).
The two Raman pulses share a laser light with frequency $\nu_2$, and the beams with frequency $\nu_1$ have different propagation directions (Extended Data Fig.~6b). 
It leads to a finite spatial modulation after the interference, enhancing the visibility of the phase dislocation from the vortices.

After the Raman interference, the density distribution $n_2(\textbf{r})$ in the $\lvert F=2\rangle$ state becomes,
\begin{equation}
    \begin{split}
 n_2(\textbf{r}) &= \frac{1}{4} \Big[n_1(\textbf{r})+n_1(\textbf{r}+\Delta \textbf{r})+ \\
&     {\psi}_1^{*}(\textbf{r})
    {\psi}_1(\textbf{r}+ \Delta \textbf{r})
    e^{i\Delta \textbf{k}\cdot \textbf{r}} +
     {\psi}_1(\textbf{r})
     {\psi}_1^{*}(\textbf{r}+\Delta \textbf{r})
     e^{-i\Delta \textbf{k}\cdot \textbf{r}}\Big]
    \end{split}
\end{equation} where $n_1(\textbf{r})={\psi}_1^{*}(\textbf{r}){\psi}_1(\textbf{r}) $ and ${\psi}_1(\textbf{r})$ are the density and wavefunction of the atoms in $\lvert F=1, m_z=1\rangle$ state, respectively. 
The wavevector difference between two Raman beams, $\Delta \mathbf{k} = \mathbf{k}-\mathbf{k}^{'}$, leads to atomic displacement $\Delta \textbf{r} = 2\hbar\mathbf{k}\Delta t/M$ after the Raman pulse time $\Delta t$.

When the condensate has no phase singularity, we observed the periodic line patterns of 12.8~$\mu$m, which can be explained by the velocity difference  $\hbar \Delta k/m$.
In contrast, the two-to-one (three-to-one) fork-shaped interference patterns can be observed, which indicate $2\pi~(4\pi)$ winding around a vortex core.
In probing the eccentric fractional skyrmion, we simultaneously read out the magnetization and phase singularity in a single experiment.
To achieve both images, half of the atoms in the $\lvert 1,1\rangle$ state are transferred into the upper hyperfine state for $\textit{in situ}$ magnetization imaging, and the remnant atoms are then used for matter-wave interference. 
This measurement allows us to identify the phase defect at the position where the magnetic droplet is observed. \\

\noindent
\textbf{Theoretical Methods.}
The mean-field order parameter $\Psi({\bf r},t)$ for the two-dimensional spin-1 Bose-Einstein condensates (BECs) is given by $\Psi({\bf r},t)=\left[\Psi_{+1}({\bf r},t),\Psi_{0}({\bf r},t),\Psi_{-1}({\bf r},t)\right]^{T}$
where $\Psi_m({\bf r},t)$ is the macroscopic wave function, ${\bf r}\in\mathbb{R}^2$ is the position, and $m$ is the Zeeman number $(m=-1,0,+1)$ representing the spin direction of each component. The dynamics of the order parameter is described by the Gross-Pitaevskii equation which can be derived from the following Lagrangian density~\cite{Kawaguchi2012}:
\begin{gather}\label{GP_lag}
    \mathcal{L}=\sum_m\bigg(i\hbar\Psi_m^*\partial_t \Psi_m-\frac{\hbar^2}{2M}(\nabla\Psi_m^*)\cdot(\nabla\Psi_m)\bigg)-\mathcal{U}~,\\
    \nonumber \mathcal{U}=\frac{c_n}{2}n^2-\mu n+\frac{c_s}{2}{\bf s}^2-ps_z+\Psi^{\dagger}(q\check{\sigma}_z^2+\check{U})\Psi~ \, .
\end{gather}
Here, the particle density $n$, the spin density ${\bf s}$ and the external potential $\check{U}$ are given by 
\begin{gather}
    n=\Psi^{\dagger}\Psi=\sum_m\Psi^*_m\Psi_m~,\\
    {\bf s}=(s_x, s_y, s_z)^T=\Psi^{\dagger}\check{{\boldsymbol \sigma}}\Psi=\sum_{mm'}\Psi_m^*(\check{{\boldsymbol \sigma}})_{mm'}\Psi_{m'}~,\\
    \check{U}=U_{\rm ext}\check{I}
\end{gather}
with the spin-1 matrices $\check{{\boldsymbol \sigma}}=(\check{\sigma}_x, \check{\sigma}_y, \check{\sigma}_z)^T$,
The external potential for $\Psi_m$ is given by  $U_{\rm ext}=\frac{1}{2}M(\omega_x^2x^2+\omega_y^2y^2)$,
$\mu$, $p$ and $q$ are the chemical potential, the linear Zeeman coefficient and the quadratic Zeeman coefficient respectively; $c_n$ and $c_s$ represent the density-dependent and the spin-dependent interaction, respectively. Note that the spin-dependent interaction gives rise to the coupling between different Zeeman components. It is known that the spin-1 BEC has four phases depending on $q$ and $c_s$~\cite{Kawaguchi2010}. In this work, $q$ and $c_s$ are negative, and, therefore, our system is in an easy-axis ferromagnetic phase.

For numerical computation, it is convenient to rescale the time variable $t$, the spatial coordinates $(x,y)$ and the wave function $\Psi_m$ to 
\begin{gather*}
    (x,y)\rightarrow(x\xi,y\xi)~,\\
    t\rightarrow t\tau~,\\
    \Psi_m\rightarrow\Psi_m\sqrt{n_F} \, ,
\end{gather*} 
where $\xi=\frac{\hbar}{\sqrt{M(\mu-q)}}~,\quad\tau=\frac{\hbar}{\mu-q}~,\quad n_F=\frac{\mu-q}{c_n+c_s}$
are natural units for length, time, and density, respectively. After the rescaling, $t, x, y$ and $\Psi_m$ are dimensionless variables.

Simulation is performed in the following sequences. First, we prepare a domain wall along the $x$-axis using numerical relaxation, the details of which are as follows. We choose the arrested Newton's flow(ANF) method~\cite{Gudnason2020} as the numerical relaxation method, which can be summarized as
\begin{gather}
    \frac{d^2\Psi_m}{dt'^2}=-\frac{\delta E[\Psi]}{\delta\Psi_m^*}~.
\end{gather}
Here, $E[\Psi]$ is the energy functional of the system, $\delta$ represents the functional derivative and $t'$ is the fictitious time. For the initial condition, we set $\Psi_{+1}/\sqrt{n_F}=1$ for upper 1/3 area, $\Psi_0/\sqrt{n_F}=1$ for middle 1/3 area and $\Psi_{-1}/\sqrt{n_F}=1$ for lower 1/3 area (upper, middle, lower concerning the vertical $y$-axis), and also set $d\Psi_m/dt'=0$. After every time evolution $t\rightarrow t+\Delta t$, we check energy as a function of time $E(t)=E[\Psi({\bf r},t)]$. If energy is increased, $E(t)<E(t+\Delta t)$, set $d\Psi_m/dt'|_{t+\Delta t}=0$ to remove the kinetic energy and restart the relaxation. The process is repeated until $\delta E=0$ within tolerance (we used $10^{-6}$). After the numerical relaxation, we obtain a ferromagnetic domain wall state aligned along the $x$-axis. We employ an anisotropic harmonic external potential, which yields an elliptic BEC with a domain wall in an equilibrium state. In addition, to mimic random fluctuations that are present in the experimental platform, we can consider the random fluctuations of the following degrees of freedom in numerical simulations: The position of the domain wall, the angle of the spin vector in the domain wall, and the overall noise. The fluctuation of the domain wall position deforms the BEC along the $y$-axis. The overall noise is implemented by adding the random number $r=de^{i\theta}$ where $r\in[0, 0.1]$ and $\theta\in[0, 2\pi)$ to the discretized $\Psi$. If domain wall position fluctuation is very small, we can give small modulation to the domain wall while minimizing the distortion of the elliptical shape of the BEC system.

 Second, we apply the magnetic field with finite gradient ${\bf B}=B_s'x\hat{z}$ where $B_s'=9.0~\text{mG/cm}$ for $t=0\sim80$ ms and simulate the time evolution of the wave function. The time evolution of the wave function is governed by the Gross-Pitaevskii equation which is derived from the Lagrangian of Eq.~(\ref{GP_lag}). We used the fixed-point iteration of the Crank-Nicolson method~\cite{ICN} with Neumann boundary condition $\frac{\partial\Psi}{\partial x}=\frac{\partial\Psi}{\partial y}=0$ at the system boundary. The application of a magnetic-field-gradient induces a non-zero relative velocity between the $m=+1$ component and $m=-1$ component and the resultant shear causes the deformation of the domain wall. This eventually leads to the Kelvin-Helmholtz instability and the generation of conventional skyrmions and eccentric skyrmions from the flutter finger patterns. See Extended Data Fig.~4a-d for the examples of simulated states.

To analyze the Kelvin-Helmholtz instability observed in numerical simulations, we perform the Fourier transform of the domain-wall configuration in the following steps. First, we extract points in the domain wall and conduct the linear fitting to obtain a straight line along the domain wall. Next, we calculate the vertical distances between the obtained straight line and the domain-wall points. Finally, we represent these values as the function of coordinate along the straight line, perform the Fourier transform on this function, and obtain the wavevector $k_{\text{max}}$ which has the maximum Fourier-transform amplitude. We repeat this process over 10 samples and calculate the average value of $k_{\text{max}}$ for each gradient field $B_s'=$ 0, 1.88, 4.50, 6.75 and 9 mG/cm. The Fourier transform is performed with SciPy library~\cite{SciPy}.

As observed in experiments shown in Fig.~3b of the main text, we also observed the pair generation of eccentric fractional skyrmions (EFSs) in the numerical simulation. Extended Data Fig.~4e-j shows one example of such instances, where an integer skyrmion with antiferromagnetic spin singularities (i.e., points where the spin density is zero) split into two EFSs. It is reminiscent of the dissociation of a doubly charged vortex into two singly charged vortices~\cite{Shin2004,Weiss2019}. In the longitudinal magnetization $F_z$, one droplet is emitted from the domain wall as shown in Extended Data Fig.~4f. After 25~ms, this droplet is split into 2 EFSs (Extended Data Fig.~4h-j), each of which possesses one spin singular point. The singular points in these droplets exhibit a ``C" shape in the density profile of the $m=0$ component as shown in Extended Data Fig.~4g and j.

\begin{figure*}[t]
\centering
\includegraphics[width=0.95\linewidth]{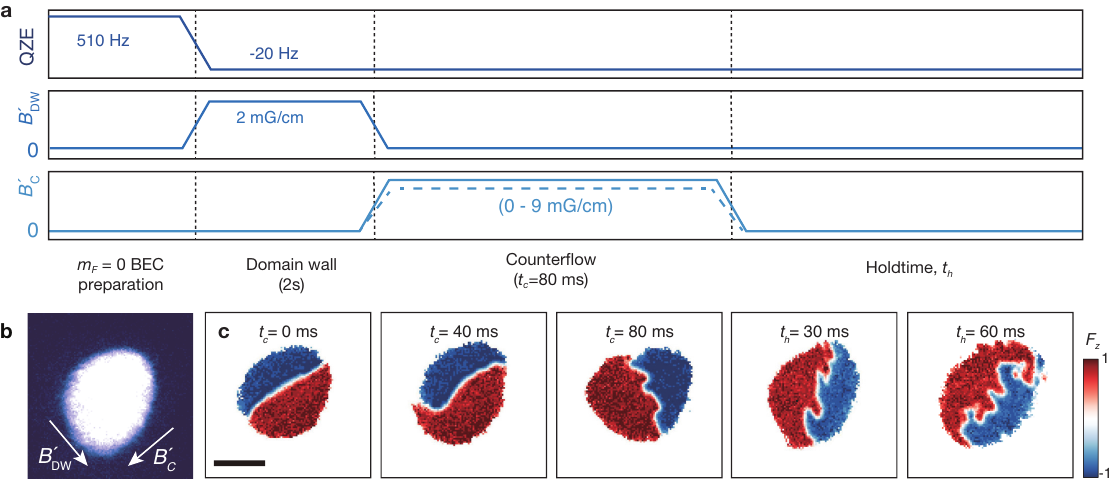}
\caption{\textbf{Experiments of the quantum Kelvin-Helmholtz instability.}
\textbf{a,} The experimental sequence consists of four different time zones: degenerate gas production, domain wall preparation, counterflow injection, and hold time without field gradient. 
\textbf{b,} Optical density image of the polar condensate before quenching the quadratic Zeeman energy (QZE). 
A field gradient $B_{\rm DW}'$ is applied along the $y$-axis to prepare a well-aligned single-domain wall state during the coarsening dynamics. The counterflow in the interface layer is generated by different gradient fields $B_c'$ along the $x$-axis. 
\textbf{c,} The snapshot images of longitudinal magnetization distribution $F_z$ during the experimental sequences. 
The gradient pulse also introduces a rotational motion of the DW boundary.
After a hold time of ($t_h=30$~ms), flutter-finger patterns are well developed.
The scale bar corresponds to 100~$\mu{}$m. 
}
\label{figS_Seq}
\end{figure*}

\newpage 

\begin{figure*}[t]
\centering
\includegraphics[width=0.75\linewidth]{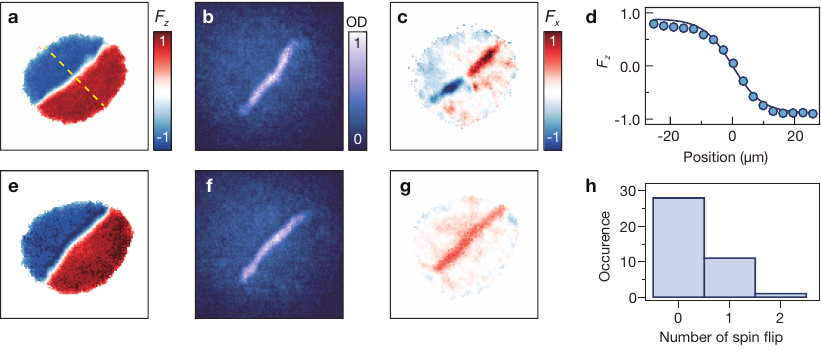}
\caption{\textbf{Broken-axis core domain wall.}
\textbf{a,} Longitudinal magnetization density $F_z$ of the broken-axis (BA) core domain wall state at $q/h=-20$~Hz.
\textbf{b,} Density profile the spin $\ket{0}$  state, and 
\textbf{c,} its transverse magnetization density $F_x$ of the BA-core domain wall.The transverse magnetization $F_x$ turns its sign at the Bloch line. The Bloch line is represented as the opposite spin vector in the transverse magnetization. 
\textbf{d,} Central cross-section profile of the longitudinal magnetization (yellow dashed line in a).
\textbf{e-g,} BA-core domain wall without the Bloch line.The transverse spin vector points in the same direction. 
\textbf{h,} The histogram of the spin-flip occurrence at the interfaces of the domain wall. 
}
\label{figS_BAcore}
\end{figure*}

\newpage

\begin{figure*}[h]
\centering
\includegraphics[width=0.7\linewidth]{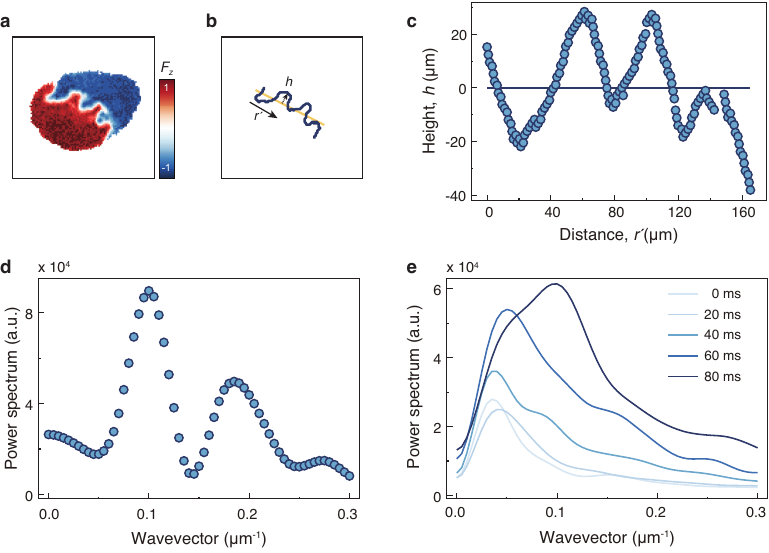}
\caption{
\textbf{Data analysis for Kelvin-Helmholtz instability.}
\textbf{a,} Exemplary image of longitudinal magnetization $F_z$, displaying the flutter-finger pattern.
\textbf{b,} Trajectory of domain wall in ({A}), which is obtained by finding the points where $F_z(\mathbf{r}_{\rm DW})= 0$.
\textbf{c,} The height difference from the linear fit results as a function of the projected distance along the fitted line.
\textbf{d,} Spline interpolated power spectrum of the interface modulation.
\textbf{e,} Averaged power spectrum of domain wall waves for various pulse durations.
}
\label{figS_Analysis}
\end{figure*}

\newpage 

\begin{figure*}[t]
\centering
\includegraphics[width=0.85\linewidth]{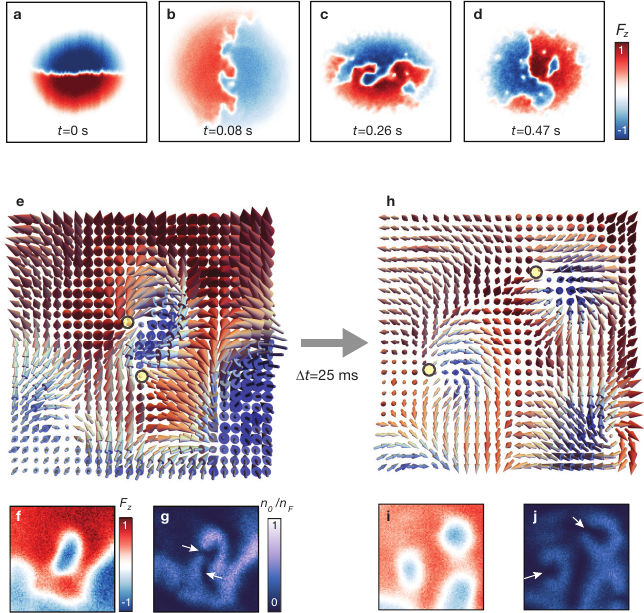}
\caption{
\textbf{Numerical simulation of the Kelvin-Helmholtz instability and EFS pair generation.}
\textbf{a-d,}  Longitudinal spin density $F_z$ at different hold times $t$. Because of the harmonic potential, the domain wall axis rotates. 
\textbf{b,}  The KHI is represented as the Flutter-Finger pattern at $t=0.08$~s. ({c,d}) At later time, magnetic droplets are emitted from the fingertip. 
\textbf{e,} The Zoomed plot for local spin vector, \textbf{f,} longitudinal spin density $F_z$, and \textbf{g,} density distribution $n_0$ of spin $\ket{0}$ state at $t=0.11$~s. 
The yellow circles indicate the region where the BEC is in the local antiferromagnetic phase with zero spin density.
\textbf{h,} After 25~ms, two EFSs with the same spin vorticity are generated from the skyrmion-like spin texture.
As a result, \textbf{i,} two magnetic droplets and \textbf{j,} two ``C" shapes are observed in the $F_z$ and $n_0$, respectively.
The spin singular points in {e} and {h} are present at the same locations as the dark nodes of the ``C" shape in g and i, respectively.
}
\label{FigBEC_sim}
\end{figure*}

\newpage

\begin{figure*}
\centering
\includegraphics[width=0.6\linewidth]{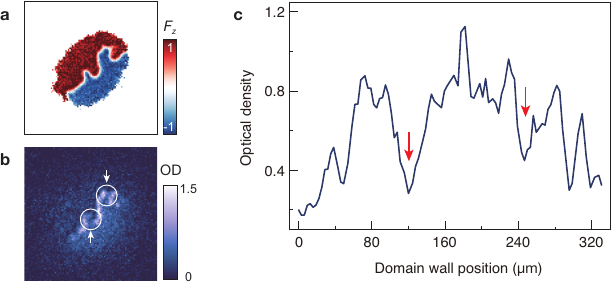}
\caption{\textbf{Spin singularity along the domain wall.} 
\textbf{a,} Longitudinal magnetization after applying gradient pulse
\textbf{b,} Absorption image of the spin $\ket{0}$ state. Anti-ferromagnetic spin singular point is highlighted by the white circle.
\textbf{c,} Density profile of the along the domain wall. 
Density dips (red arrows) denote the singular point.
}
\label{figS_DWsing}
\end{figure*}

\begin{figure*}
\centering
\includegraphics[width=0.75\linewidth]{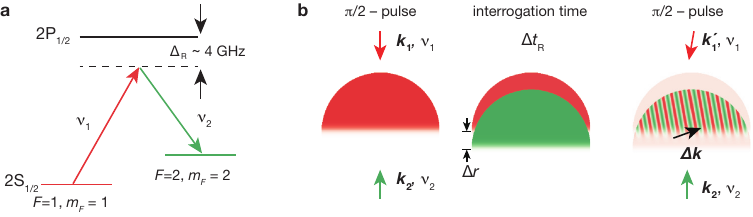}
\caption{\textbf{Matter-wave interference using two-photon Raman transition.}
\textbf{a,} Schematic diagram of Raman transition. 
All Raman beams are 4~GHz red detuned to $D_1$ transition line, and the frequency difference is set to $\Delta \nu=\nu_1-\nu_2=803.8$~MHz, corresponding to the hyperfine splitting between the $\lvert 1, 1 \rangle$ and $\lvert 2, 2 \rangle$ states. 
The laser beams have lin$\perp$lin configuration and their beam waist are 1~mm, larger than the condensate diameter ($\sim$ 300~$\mu$m).
\textbf{b,} Experimental sequence for matter-wave interference. 
We first apply the $\pi/2$ pulse of the Raman transition with a counterpropagating configuration $(\mathbf{k},\mathbf{-k})$. 
Then, the copy of condensate in $\lvert 2, 2 \rangle$ state can move along the direction of Raman lasers with a photon recoil momentum $2\hbar\mathbf{k}$.
Having a time interval of $\Delta t_R=200~\mu{}$s between the two Raman pulses, the condensates in each hyperfine spin state have $\Delta r=34~\mu${m} of spatial displacement.
The sequence ends with applying the second $\pi/2$ pulse of Raman beams with its mutual angle of 3$^\circ$.
The difference in photon recoil momentum from the Raman pulse results in the spatial modulation in the atomic density profile.
}
\label{FigS_Raman}
\end{figure*}

\begin{figure*}
\centering
\includegraphics[width=0.65\linewidth]{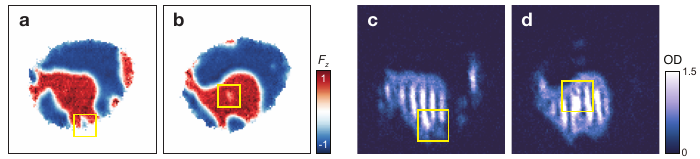}
\caption{
    \textbf{Eccentric fractional skyrmion decay.}
 \textbf{a,b,} Longitudinal magnetization and \textbf{c,d,} Raman interference images with EFS after 2~s of hold time. 
    The EFS can be lost by drifting outside of the condensate (a,c) or decay to a vortex without a magnetic core (b,d).
}
\label{FigS_decay}
\end{figure*}

\clearpage
\newpage

\end{document}